\documentclass[twoside]{article}
\usepackage{fleqn,espcrc2,amsmath,amssymb}

\usepackage{graphicx}

\title{Fixed Point SU(3) Gauge Actions:\ Scaling Properties and
Glueballs \thanks{Work supported in part by Schweizerischer
Nationalfonds} \thanks{Talk presented by U. Wenger} }

\author{Ferenc Niedermayer\address[MCSD]{Institute of Theoretical Physics, 
       University of Bern, Sidlerstrasse 5, CH-3012 Bern, Switzerland},
        Philipp R\"ufenacht\addressmark[MCSD] and
        Urs Wenger\address{Theoretical Physics, Oxford University, 1
        Keble Road, Oxford OX1 3NP, United Kingdom}}
       
\begin{document}

\begin{abstract}
We present a new parametrization of a SU(3) fixed point (FP) gauge
    action using smeared ("fat") gauge links. We report on the
    scaling behaviour of the FP action on coarse lattices by means of
    the static quark-antiquark potential, the hadronic scale $r_0$,
    the string tension $\sigma$ and the critical temperature $T_c$ of
    the deconfining phase transition.  In addition, we investigate the
    low lying glueball masses where we observe no scaling violations
    within the statistical errors.
\end{abstract}

\maketitle

\section{Introduction}
It is well known that on rather coarse lattices with lattice spacings
$a \geq$ 0.1 fm the Wilson gauge action produces quite large cut-off
effects. On the other hand, decreasing the lattice spacing below 0.1
fm becomes increasingly difficult for simulations including
fermions. Therefore using a better, i.e.~improved gauge action might
well be important. 

One possible way of constructing non-per\-tur\-ba\-ti\-ve\-ly improved
lattice gauge theories is provided by the fixed point (FP) action
approach \cite{Hasenfratz:1994sp}. FP actions are classically perfect,
i.e.~they have no lattice artefacts on classical
con\-fi\-gu\-ra\-tions. In particular, they possess exactly scale
invariant instanton solutions. Together with fermion FP actions they
form appealing formulations of non-perturbatively improved lattice QCD
\cite{Hasenfratz:1998bb}.

Here we present a new parametrization for FP gauge actions which we
apply to SU(3). The new parametrization is much more flexible and has
a richer structure than the loop parametrizations studied so far
\cite{DeGrand:1995ji,Blatter:1996ti} and
therefore allows to better reproduce the classical properties of the
theory.  In order to check the scaling behaviour of the parametrized
FP action we perform a series of simulations in which we determine the
critical temperature of the deconfining phase transition, the static
$\bar q q$ potential and glueball masses. This paper summarizes some
of the results \cite{Niedermayer:2000yx}.

\section{FP gauge actions}
We consider SU($3$) pure gauge theory in four dimensional Euclidean
space-time on a periodic lattice.
For asymptotically free theories the determination of the FP action
reduces to a saddle point problem encoded in the FP equation
\cite{Hasenfratz:1994sp}
\begin{equation}
  \label{eq:FP_equation}
  {\cal A}^{\text{FP}}(V) = \min_{\{U\}} \left\{{\cal A}^{\text{FP}}(U) + T(U,V)\right\},
\end{equation}
where ${\cal A}^{\text{FP}}$ is the FP action, $T(U,V)$ is the
blocking kernel of a renormalization group (RG) transformation and $U$
and $V$ are the gauge fields on the fine and coarse lattice, respectively.
 We use the RG transformation of ref.~\cite{Blatter:1996ti} where the
parameters of the transformation have been optimized for a short
interaction range of the FP action and improved rotational invariance.
Eq.~(\ref{eq:FP_equation}) represents an implicit equation for the FP
action which, in principle, contains infinitely many couplings. In
order to use it in MC simulations one has to find an
appropriate parametrization.

\section{The new parametrization}
The main novelty in the new parametrization of the FP gauge action is
the use of smeared (``fat'') links which are used to build simple
Wilson loop plaquettes \cite{Niedermayer:2000yx}. To build a plaquette
in the $\mu\nu$-plane from smeared links we introduce asymmetrically
smeared links: instead of summing up all staples contributing
to a smeared link we are suppressing the staples lying in the
$\mu\nu$-plane relative to those in the orthogonal planes
$\mu\lambda$, $\lambda\ne\nu$. In this way we are able to effectively
generate long and complicated loops which, however, are still very
compact.

From these ``smeared'' loops $W_{\mu\nu}^{\text{pl}}$ and from the unsmeared
standard Wilson loop $U_{\mu\nu}^{\text{pl}}$ we
construct smeared and unsmeared plaquette variables $w_{\mu\nu}$
and $u_{\mu\nu}$, respectively,
\begin{eqnarray*} 
u_{\mu\nu} & = & \text{Re Tr}(1-U_{\mu\nu}^{\text{pl}}),\\
w_{\mu\nu} & = & \text{Re Tr}(1-W_{\mu\nu}^{\text{pl}}),
\end{eqnarray*} 
which in turn are used in a mixed polynomial ansatz
for the FP action
\begin{equation}\label{eq:polynomial ansatz}
{\cal A}[U]=\frac{1}{N_c}\sum_{x,\mu<\nu}\sum_{k,l} p_{kl}\,u_{\mu\nu}(x)^k\, w_{\mu\nu}(x)^l\,.
\end{equation}
The linear parameters $p_{kl}$ and the non-linear ones entering the
smearing are determined through eq.~(\ref{eq:FP_equation}) which is
used recursively connecting coarse configurations to smoother and
smoother ones. On the smoothest configuration we use the analytically
calculable quadratic approximation to the FP action as the starting
point \cite{Niedermayer:2000yx}.

An additional novelty is the use of local information stored in the
fine configurations minimizing the r.h.s.~of
eq.~(\ref{eq:FP_equation}) for the fitting procedure. Obviously the
minimizing configurations contain much more information than just the
total value of the action. To explore this information we calculate
the derivatives of the FP action with respect to the gauge links in a
given colour direction at each site and, together with the action
values, include them in the fitting procedure for determining the
parameters.

The final parametrization \cite{Niedermayer:2000yx} obtained in this way
 contains five non-linear parameters entering
the smearing and 14 linear ones $\{p_{kl}, 0< k+l \leq 4\}$ describing
the mixed polynomial in eq.~(\ref{eq:polynomial ansatz}).

\section{Physical results}
Using the parametrized FP action we perform a large number of
simulations on coarse lattices with lattice spacings $a$ between 0.1
and 0.33 fm in order to explore the scaling behaviour of different
physical quantities. In all simulations we generate independent gauge
configurations using a mixture of Metropolis and overrelaxation
updates. Wherever it is applicable we make use of smearing techniques
and employ variational techniques to increase the overlap of wave
functions with the ground state \cite{Niedermayer:2000yx}.

\subsection{The critical temperature}
We first measure the critical temperature of the deconfining phase
transition using the Polyakov loop $L$ as an order parameter. To do
so we determine the critical couplings on lattices with temporal
extent $N_\tau = 2,3$ and 4 and various spatial volumes from the
location of the peak in the Polyakov loop susceptibility,
\begin{equation*}
  \label{eq:def_Pol_loop_susc}
  \chi_L \equiv V_\sigma 
  \left( \langle|L|^2\rangle - \langle|L|\rangle^2 \right),
  \quad V_\sigma = N_\sigma^3.
\end{equation*}
The location of the peak can be determined very precisely by a
spectral density reweighting \cite{Ferrenberg:1988yz} which combines
the data of many simulations at values of the gauge coupling near the
critical value. In order to perform the thermodynamic limit we resort
to the finite scaling behaviour linear in $V_\sigma$ as expected for a
first order phase transition.  The results from these extrapolations
are $\beta_c^{N_\tau = 2} = 2.361(1), \, \beta_c^{N_\tau = 3} =
2.680(2)$ and $\beta_c^{N_\tau = 4} = 2.927(4)$.  Finally, the
critical temperature is determined through $a(\beta_c) T_c = 1/N_\tau$.

\subsection{The static $\bar q q$ potential}
\begin{figure}[htbp]
  \begin{center}
    \includegraphics[width=7.5cm]{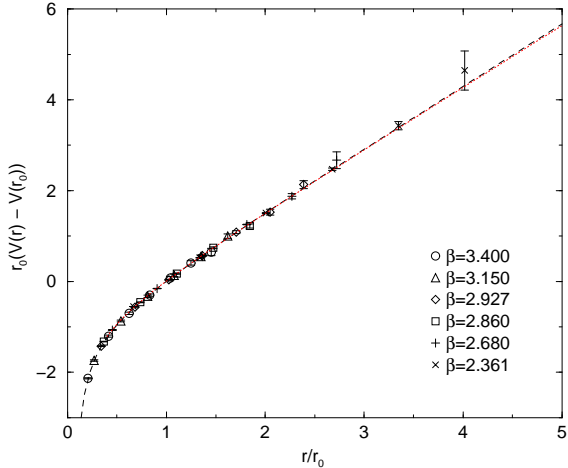}
    \caption{{}Scaling of the static $q\bar{q}$ potential $V(r)$
    expressed in terms of the hadronic scale $r_0$. }
    \label{fig:all_pot}
  \end{center}
\end{figure}

One way of examining the scaling of the parametrized FP action is by
measuring the static $q\bar{q}$ potential $V(r)$ in terms of the
hadronic scale $r_0$ and comparing the quantity $r_0\left(
V(r)-V(r_0)\right)$ versus $r/r_0$ at several values of $\beta$ (see
figure \ref{fig:all_pot}).
We would like to stress that on coarse lattices this a non-trivial
scaling test. The
quality of our measurements can be seen for example by comparing our
fit (dashed line) of the form Coulomb plus linear term with the
result (dotted line) from measurements with an anisotropic
tadpole/tree level Symanzik improved action \cite{Juge:1997nc}.

The hadronic scale used above is determined from the
force $F(r)$ between two static quarks. We use \cite{Sommer:1994ce}
\begin{equation}
  \label{eq:def_hadronic_scale}
  r_0^2 V'(r_0) = r_0^2 F(r_0) = 1.65, 
\end{equation}
where $r_0 \approx 0.49 \, \text{fm} = (395 \,
\text{MeV})^{-1}$. However, since eq.~(\ref{eq:def_hadronic_scale})
needs an interpolation of the force in between lattice sites, on
coarse lattices this definition of the scale is plagued by ambiguities
stemming from the discreteness of the lattice points.

Another physical quantity of interest is the string tension $\sigma$
which describes the asymptotic long range behaviour of the potential.  

Note, that here we do not aim at testing the rotational invariance of
the potential. For the RGT used here this has  been done at
finite temperature, however, using a different parametrization of the
FP action \cite{Blatter:1996ti}.

\subsection{Scaling behaviour}
We are now in a position to investigate the scaling behaviour of the
parametrized FP action by means of the RG invariant quantities $r_0
T_c, T_c/\sqrt{\sigma}$ and $r_0 \sqrt{\sigma}$. 
\begin{figure}[htbp]
  \begin{center}
    \includegraphics[width=7.5cm]{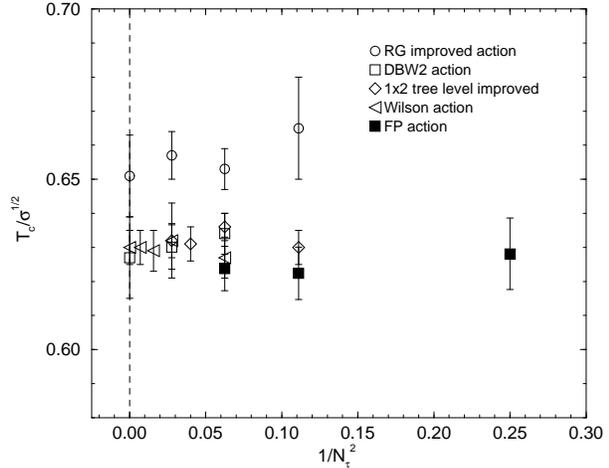}
    \caption{{}Scaling of $T_c/\sqrt{\sigma}$ for different actions.}
    \label{fig:sigmaTc}
  \end{center}
\end{figure}
We find that these quantities scale even on lattices as
coarse as $a \simeq 0.33$ fm  \cite{Niedermayer:2000yx}. As an example
we display the results for $T_c/\sqrt{\sigma}$ in figure
\ref{fig:sigmaTc} together with results obtained with different other
actions \cite{Iwasaki:1997sn,Beinlich:1997ia,Bliss:1996wy,Boyd:1996bx,Edwards:1997xf,deForcrand:1999bi}.

\subsection{Glueball masses}
\begin{figure}[htbp]
  \begin{center}
    \includegraphics[width=7.5cm]{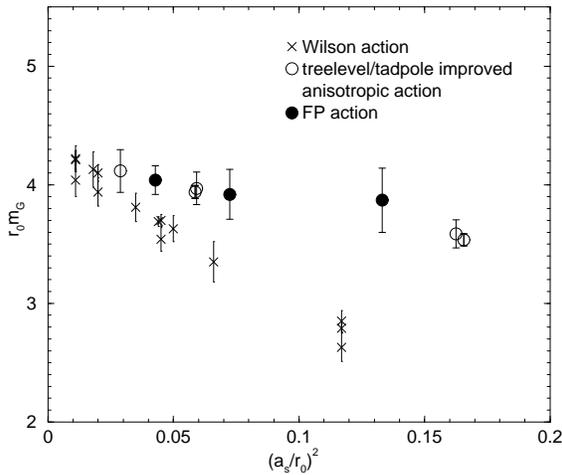}
    \caption{{}Scaling behaviour of the $0^{++}$ glueball.}
    \label{fig:0pp_final}
  \end{center}
\end{figure}

In the pure gauge glueball spectrum the lowest-lying $0^{++}$ state
shows particularly large cut-off effects when measured with the Wilson
action. It therefore provides an excellent candidate for sizing the
improvements achieved with the parametrized FP action. To do so we
perform simulations at three different lattice spacings in the range
$0.1 \, \text{fm} \leq a \leq 0.18 \, \text{fm}$ and spatial volumes
between (1.4~fm)$^3$ and (1.8~fm)$^3$. The scale is set by $r_0$ as
determined in the previous section. We measure all length eight loops
on five smearing levels in order to construct a wave function with a
large overlap to the ground state.  In figure \ref{fig:0pp_final} we
compare the results for the lowest lying $0^{++}$ glueball mass
measured with the Wilson action
\cite{Teper:1998kw,Bali:1993fb,Vaccarino:1999ku}, the anisotropic
tadpole/tree level Symanzik improved action
\cite{Morningstar:1997ff,Liu:2000ce} and the FP action.
Performing the continuum limit and using $r_0 \simeq 0.49$ fm we
obtain a mass of $m_{0^{++}} = 1627(83)$ MeV for the $0^{++}$
glueball. Finally, we also determine the $2^{++}$ glueball mass at a
lattice spacing $a = 0.1$ fm and obtain $m_{2^{++}} = 2354(95)$ MeV
\cite{Niedermayer:2000yx}.

Measuring glueball states on rather coarse lattices immediately calls
for anisotropic gauge actions. The construction of such an action
using the FP approach is under way \cite{Rufenacht:2000px}.

\section{Conclusions}
The new parametrization presented here provides a method for
approximating FP gauge actions which is very general and flexible but
still simple. It is possible to describe the FP actions accurately
enough so that their classical properties are preserved. However, it
is necessary to check the range of validity of the parametrized FP
action. In view of future applications of the action in
connection with a parametrized FP Dirac operator
\cite{Hasenfratz:2000qb} this is of crucial importance. In addition,
one has to assure that no pathologies are introduced through the
parametrization.  For the quantities and lattice spacings investigated
so far we find that this is indeed the case. In particular, we find
that these physical quantities scale even on lattices as coarse as
$a=0.33$ fm.

{\bf Acknowledgements:} We would like to thank Peter Hasenfratz for useful suggestions and discussions.

\end{document}